\documentclass[dvips]{arxstspdf}
\usepackage{graphics,flushend}

\volume{25}
\issue{2}
\pubyear{2010}
\firstpage{191}
\lastpage{201}
\doi{10.1214/10-STS334}

\makeatletter

\def\citep#1{(\cite{#1})}
\newproclaim{Property}{Property}
\makeatother

\begin{document}
\begin{frontmatter}

\title{Stochastic Approximation and Modern Model-Based Designs for  Dose-Finding Clinical Trials}
\runtitle{Stochastic Approximation and  Modern Dose-Finding}

\begin{aug}
\author{\fnms{Ying Kuen} \snm{Cheung}\corref{}\ead[label=e1]{yc632@columbia.edu}}
\runauthor{Y. K. Cheung}

\address{Ying Kuen Cheung is Associate Professor, Department of Biostatistics,
Columbia University, New York, 10032 New York, USA \printead{e1}.}

\end{aug}

\begin{abstract}
In 1951 Robbins and Monro published the seminal article on  stochastic approximation and made
a specific reference to its application to the ``estimation of a quantal using response,
nonresponse data.''  Since the 1990s, statistical methodology for dose-finding studies
has grown into an active area of research.  The dose-finding problem is at its
core a  percentile estimation problem and  is in line with what the Robbins--Monro method sets out to solve.
In this light, it is quite surprising that the dose-finding literature has developed rather
independently of the older  stochastic approximation literature.
The fact that stochastic approximation has seldom been used
in actual clinical studies stands  in stark contrast with
its constant application  in engineering and finance.
In this article, I explore similarities and differences between
the  dose-finding  and the stochastic approximation literatures.
This review also sheds light on the present and future relevance
of stochastic approximation to dose-finding clinical trials.
Such connections will in turn steer  dose-finding methodology on a rigorous course
 and extend its ability to handle increasingly complex clinical situations.
\end{abstract}


\end{frontmatter}

\section{Introduction}

Dose-finding in phase I clinical trials is typically  formulated as
estimating a prespecified percentile of a dose-toxicity curve.  That is, the objective is
to identify a dose $\theta$ such that $\pi(\theta) = p$, or equivalently,
\begin{equation}\label{eq:obj0}
\theta = \pi^{-1}(p),
\end{equation}
where $\pi(x)$ is the probability of toxicity
at dose $x$ and is assumed continuous and increasing in $x$.
Percentile estimation, often seen in bioassay, is a well-studied problem
for which statisticians have an extensive set of tools; see the books by  \citet{finney} and \citet{morgan}.
There are, however, two practical aspects of clinical
studies that distinguish phase I dose-finding from the classical bioassay problem.
 First,  the experimental units are humans. An implication is that
the subjects should be treated sequentially with respect to some ethical constraints
(e.g., Section~\ref{sec:cohere}).
 As such, dose-finding is as much a {\em design} problem as an analysis problem.
 Second, the actual doses administered to the subjects are confined to a discrete panel of levels,
denoted by $\{d_1, \ldots, d_K\}$, with $\pi(d_1) < \cdots < \pi(d_K)$.
Therefore, it is  possible  that $\pi(d_k) \neq p$ for all $k$, and
the working objective then is to identify the dose
\begin{equation}\label{eq:obj}
\nu = \arg \min_{d_k} | \pi(d_k) - p |.
\end{equation}
Apparently, the continuous dose-finding objective $\theta$ and the discrete objective $\nu$ are close to
each other.  However, in this article, we will see that  discretized versions of
methods developed for $\theta$ are not necessarily good solutions for $\nu$.
This special section of {\em Statistical Science} also
consists of four other articles that review some benchmarks in the recent development of the so-called
model-based methods for dose-finding studies.
In a nutshell, a model-based method makes
dose decisions based on the explicit use of a dose-toxicity model.  That is, the toxicity
probability at dose $x$, $\pi(x)$,
is postulated to be $F(x, \phi_0)$ for some true parameter value $\phi_0$.
This is
in contrast to the class of algorithm-based designs whereby a set of dose-escalation rules
are prespecified for any given dose without regard to the observations at the other doses.
Section~\ref{sec:df} of this article will present a brief history of the development of the
modern dose-finding methods and define the scope of this special issue.

In addition, this article complements the other articles in two ways.  First, it
consolidates the key theoretical dose-finding criteria that are otherwise
scattered in the literature (Section~\ref{sec:theory}).
Second, it compares and contrasts
the  dose-finding literature with the large  literature on
stochastic approximation  (Section~\ref{sec:past});
 the former primarily addresses the discrete objective $\nu$, whereas the
latter deals with  $\theta$.
While this literature synthesis is of intellectual interest, it also sheds light
on how we may tailor the well-studied stochastic approximation method
to meet the practical needs in dose-finding studies (Section~\ref{sec:practice}).
Section~\ref{sec:future} will end this article with some  future directions in
dose-finding methodology.\vspace{-2pt}

\section{Modern Dose-Finding Methods}\label{sec:df}\vspace{-1pt}

\subsection{A Brief History}\vspace{-1pt}

This article uses the work of \citet{storer87} as a historical line to define the modern statist\-ical literature
of dose-finding.  Little  discussion and formal formulation of the dose-finding problem existed
in the pre-1987 statistical literature; an exception was the article by \citet{anbar84}.
 While dose-finding~in~\mbox{cancer} trials was discussed
as early as in the~1960s~in~the~bio\-medical communities,
a well-defined quantitative objective such as (\ref{eq:obj}) was absent in the communications; see the work of
\citet{sch65} and \citet{geller84}, for example. The article by
\citet{storer87}  is the earliest reference,
to the best of my know\-ledge, that engages the clinical readership with the
idea of percentile estimation.
The authors point out the arbitrary estimation
properties associated
with the traditional 3$+$3 algorithm used in actual dose-finding studies in cancer patients.
The 3$+$3 algorithm identifies the so-called maximum tolerated dose (MTD) using
the following dose-escalation rules after enrolling every group of three subjects: let
$x_j$ denote the dose given to the $j$th group of subjects and suppose $x_j =d_k$; then\vspace{-2pt}
\begin{equation}\label{eq:3+3}
x_{j+1} =
\cases{ d_{k+1}, &\mbox{if $z_k/n_k < 0.33$,}
\cr
d_k, &\mbox{if $z_k = 1$ and $n_k=3$,}
\cr
d_{k-1}, &\mbox{if $z_k \geq 2$,}\vspace{-2pt}
}
\end{equation}

\noindent
where $n_k$ and $z_k$ respectively denote the {\em cumulative} sample size and number
of toxicities at dose $d_k$.
The trial will be terminated once a de-escalation occurs, and the next lower dose will be called the MTD.
In the sequel,
\citet{storer89} deduced from the 3$+$3 algorithm (\ref{eq:3+3}) that a cancer dose-finding study aims to
estimate the 33rd percentile (i.e., $p=0.33$).  While it has now emerged that the target is likely  lower
than the 33rd percentile with $p$ being between 0.16 and 0.25, their work has
shaped the subsequent
development of dose-finding methods in both the statistical and biomedical literatures, and the
MTD  has since been  defined invariably
as a dose associated with a prespecified toxicity probability $p$.

O'Quigley, Pepe and Fisher (\citeyear{oq90}) proposed the continual reassessment method (CRM) in 1990. The CRM
is the first model-based method in the modern dose-finding literature.  The main idea of the method is
to treat the next subject or group of subjects at the dose with toxicity probability
estimated to be closest to the target $p$.  Precisely, suppose we have observations from the
first $j$ groups of subjects and compute the posterior mean $\hat \phi_j$ of $\phi$ given
these observations.
Then the next group of subjects will be treated at
\begin{equation}\label{eq:crm}
x_{j+1} = \arg \min_{d_k} | F(d_k, \hat{\phi}_j) - p |.
\end{equation}
A  similar idea  is adopted in most model-based designs proposed since 1990.   One  example is
 the escalation with overdose control (EWOC) by \citet{babb98}, who applied the
continual reassessment notion
but estimated the MTD with respect to
an asymmetric loss function which places heavier penalties on overdosing than underdosing.
\citet{oq-sts} and \citet{rogatko-sts} in this special issue
 review the CRM and the EWOC and their respective extensions.
Another CRM-like design is the curve-free method  by
\citet{ge2000} who estimated the dose-toxicity
curve using a Bayesian nonparametric\break method
in an attempt to avoid bias due to model misspecification.
\citet{leung01}
proposed an
analogous frequentist version that uses isotonic regression for estimation.
  Other model-based designs
include the Bayesian decision-theoretic design\break \citep{whitehead95},
 the logistic dose-ranging strategy \citep{murphy97}, and\break
Bayesian $c$-optimal design \citep{haines03}.

The late 1990s saw  an increasing interest in algo\-rithm-based designs.
\citet{durham97} proposed a biased coin design by which
the dose for the next subject is reduced if the current subject has a toxic outcome, and the dose is
escalated with a probability $p/(1-p)$  otherwise.  The biased coin design is a randomized
version of the \citet{dixon48} up-and-down design.
Motivated by its similarity to the traditional design,
\citet{cheung07} studied a class of stepwise procedures that includes (\ref{eq:3+3})  as
a special case.
Yet another algorithm-based method was proposed by \citet{ji07}
who made interim decisions based on
the posterior toxicity probability interval associated with each dose.
The impetus for these algorithm-based designs
is simplicity:
the decision rules can be charted prior to the trial, so that the clinical investigators know exactly
how doses will be assigned based on the observed outcomes.

In order to make dose-finding techniques relevant to clinical practice,
statisticians have responded  to the realistically complicated clinical situations such as
time-to-toxicity endpoints \citep{cheung00} and  combination treatments \citep{thall03}.
While the core  dose-finding objective remains a percentile estimation problem, the complexity
of dose-finding methods has grown rapidly in the literature, with most innovations taking
the model-based approach. \citet{thall-sts} in this special issue will review the major
development of these complex designs.

Most (model-based) designs in the literature up to this point take the myopic approach by which the dose assignment
is optimized with respect to the next immediate subject without regard to the future subjects.
\citet{bartroff}
 in this issue break away from this direction and propose a model-based method from an adaptive
control perspective.
While this work attempts to solve a specific Bayesian optimization problem, it also sets
a new direction in the modern dose-finding techniques;
see Section~\ref{sec:past} of this article.

\subsection{Why Model-Based Now}

A model-based design allows borrowing strength from information across doses.
This characteristic appeals to
statisticians and clinicians alike, especially because
of the typically small-to-moderate sample sizes seen in early-phase clinical studies.
As clinicians begin to appreciate the crucial role of   dose-finding
in the entire drug development program and the value of statistical inputs to
reconcile the ethical and
research aspects in early-phase trials, their discussions have revolved around model-based
innovations such as the CRM \citep{ratain93} and the EWOC \citep{eisen}.  The increasing number of
applications in actual trials \citep{tite-jco} indicates the clinical
awareness and readiness for these model-based methods.

When compared to the simplicity of algorithm-based methods, the model-based approaches are computationally
complex and require special programming before and during the implementation of a trial.
Thanks to the advances of computing algorithms (e.g., Markov chain Monte Carlo) and computer
technology, however, trial planning with extensive simulation has become feasible.
This being the case, a full-scale  dynamic programming can still stretch the computing resource;
see the article by  \citet{bartroff} for some comparison of computational times.
In addition, statistician-friendly software has become increasingly available for the planning
and execution of these model-based designs, for example, the \texttt{dfcrm} package in R \citep{cheung-r}.
These indicators of computational maturity transform the model-based designs into practical tools for
dose-finding trials.

Finally,
the development of dose-finding theory and dose-response models
in the past two decades
lends scientific rigor to the complexity of the model-based methods.
Indeed, the goal of this special issue is to review the theoretical and modeling progress
made in the modern dose-finding literature, and\break thereby demonstrate the full promise,
and perhaps challenges, of the model-based methods.

\subsection{Some Theoretical Criteria}\label{sec:theory}

In a typical dose-finding trial, subjects are enrolled in small groups of size $m \geq 1$.
The enrollment plan is said to be fully sequential when $m=1$.  Let
$x_i$ denote the dose given to the $i$th group of subject(s).
Thus, the sequence $\{ x_i \}$ forms the design of a dose-finding study.  As
most dose-finding methods are outcome-adaptive, each design point $x_i$ is random and depends on
the previous observation history.  Evaluation of a dose-finding method therefore involves the
study of its design space with respect to some ethical and estimation criteria.
 This section will review some key dose-finding criteria including
coherence, rigidity, indifference intervals, and unbiasedness.

\subsubsection{Coherence}\label{sec:cohere}

First, consider fully sequential trials with $m=1$, so that each human subject is an experimental
 unit.
An ethical principle, coined coherence by \citet{cheung05},
dictates that no escalation should take place for the next enrolled subject if the current
subject experiences some toxicity, and that dose reduction for the next subject is not appropriate
if the current subject has no sign of toxicity.
 Precisely, let $Y_i$ denote the toxicity outcome of the $i$th subject.
An escalation for the
subject is said to be coherent only when $Y_{i-1} = 0$; likewise, a de-escalation is coherent
only when $Y_{i-1}=1$.  Extending the notion of coherence for each move, one can naturally
define coherence as a property of a dose-finding method:

\begin{Property}[(Coherence)]\label{pr1} A dose-finding design $\mathcal{D}$
is said to be coherent in escalation if with probability~1
\begin{equation}\label{eq:cohereE}
P_{\mathcal{D}} ( U_i > 0 | Y_{i-1}=1 )= 0
\end{equation}
for all $i$, where $U_i = x_i - x_{i-1}$ is the dose increment from subject $i-1$ to $i$,
and $P_{\mathcal{D}}(\cdot)$ denotes probability computed under the design $\mathcal{D}$.
Analogously, the design is said to be coherent in de-escalation if with probability 1
\begin{equation}\label{eq:cohereD}
P_{\mathcal{D}} ( U_i < 0 | Y_{i-1}=0 )= 0
\end{equation}
for all $i$.
\end{Property}

 It is important to note that coherence  is motivated by ethical concerns,
and hence may not correspond to efficient  estimation of the dose-toxicity curve.
For example, in bioassay, an efficient design obtained by sequentially maximizing some
function of the information may induce incoherent moves, and thus is not appropriate for human trials;
see the work of \citet{mcleish}, for example.

An algorithm-based design can explicitly incorporate dose decision rules that respect
the coherence principles; see the biased coin design.  For a model-based design, on the other hand,
it is not immediately clear whether coherence necessarily holds. There are three general ways to
ensure coherence in practice.  First, one could adopt model-based methods that have been proven
coherent analytically.  This includes the one-stage Bayesian CRM.
Second, one could take a numerical approach.
Let $N$ denote the sample size of a trial.
 Then the design space is completely generated by the first $N-1$ binary
toxicity
observations, and thus consists of  $2^{N-1}$ possible design outcomes.
 Therefore, one could establish coherence (for a given $N$)
by enumerating all possible  outcomes and verifying that there is no incoherent move.
In some cases, the number of outcomes can be immensely reduced to the order of $N$; see Theorem 1
in the article by \citet{cheung05}.  Third, one could  enforce coherence by
restriction when the model-based dose assignment is incoherent.
 Applying coherence restrictions is common
in practice \citep{faries} and is the most straightforward approach for complex designs.
On the other hand, the restricted moves need to be examined carefully lest they should cause an
 incompatibility problem as defined by \citet{cheung05}.

In practice, the enrollment plan is often small-group sequential, that is, $m > 1$, in order
to reduce the number of interim decisions and hence trial duration.  In this case, each group
of subjects may be viewed as an experimental unit.  A generalized version of Property~\ref{pr1} can be
stated as:

\renewcommand{\theProperty}{\arabic{Property}$'$}
\setcounter{Property}{0}
\begin{Property}[(Group coherence)]\label{pr1'} A dose-finding design $\mathcal{D}$
is said to be group coherent in escalation if with probability 1
\begin{equation}\label{eq:gcohereE}
P_{\mathcal{D}} ( U_i > 0 | \bar Y_{i-1} \geq p )= 0
\end{equation}
for all $i$, where $U_i = x_i - x_{i-1}$ now denotes the dose increment from group $i-1$ to $i$
and $\bar Y_{i-1}$ is the observed proportion of toxicities in group $i-1$.
Analogously, the design is said to be group coherent in de-escalation if with probability 1
\begin{equation}\label{eq:gcohereD}
P_{\mathcal{D}} ( U_i < 0 | \bar Y_{i-1} \leq p )= 0
\end{equation}
for all $i$.
\end{Property}

It is easy to see that (\ref{eq:gcohereE}) and (\ref{eq:gcohereD}) reduce to
(\ref{eq:cohereE}) and (\ref{eq:cohereD}), respectively, when $m=1$ for $p \in (0,1)$.

\subsubsection{Rigidity and sensitivity}
A design sequence $\{ x_n \}$ is strongly consistent for $\theta$
if $x_n \rightarrow \theta$ with probability 1.  For trials allowing only a discrete number
of test doses as in (\ref{eq:obj}),
consistency means $x_n = \nu$ eventually with probability 1.
Consistency, a desirable statistical
property in general, has an ethical connotation in dose-finding studies
because it implies all subjects enrolled after a certain time point will be
treated at $\nu$, which is the desired dose.

\renewcommand{\theProperty}{\arabic{Property}}
\setcounter{Property}{1}
\begin{Property}[(Rigidity)]\label{pr2}  A dose-finding design $\mathcal{D}$ is said to be rigid
if for every
 $0 < p_L < \pi(\nu)  < p_U < 1$ and all $n \geq 1$,
\[
P_{\mathcal{D}} \{ x_n \in \mathcal{I}_{\pi}(p_L,p_U)  \} < 1,
\]
where $\mathcal{I}_{\pi}(p_L,p_U) = \{ x\dvtx p_L \leq \pi(x) \leq p_U \}$.
\end{Property}

It is easy to see that  consistency excludes the rigidity problem.  In other words,
Property~\ref{pr2} implies that a design is inconsistent.
In particular, rigidity occurs when a CRM-like procedure is applied in conjunction with
nonparametric estimation.  Hence, such a nonparametric design is inconsistent.
This is quite interesting and somewhat counterintuitive, because
nonparametric estimation is introduced with an intention to remove bias and to enhance
the\break prospect of consistency.

 To illustrate, consider
a design that starts at dose level 1, enrolls subjects in groups of
size $m=2$, and assigns the next group at
$\arg \min_k | \tilde{p}_k - p |$ where $\tilde{p}_k$ is an estimate of
$\pi(d_k)$ based on isotonic regression, and the target is $p = 0.20$.
Now suppose that none of the subjects in the first group has a toxic outcome.
Then suppose the second group enters the trial at dose level 2, with one of the two experiencing toxicity.
Based on these observations, the isotonic estimates are
$\tilde{p}_1 = 0.00$ and $\tilde{p}_2 = 0.50$, which bring the trial back to
dose level 1. From this point on, because there is no parametric extrapolation to affect
the estimation of $\pi (d_2)$ by the data collected at $d_1$, the isotonic estimate $\tilde p_2$
will be no smaller than 0.50 regardless of what happens at $d_1$, that is,
$| \tilde p_2 - 0.20 | \geq 0.30$.
As a result, $| \tilde p_1 - 0.20 | < 0.30 \leq | \tilde p_2 - 0.20 |$ if $\tilde p_1 \leq 0.20$.
That is, the trial will stay at dose level 1 even if
 there is a long string of nontoxic outcomes there!

This example demonstrates that nonparametric estimation and the sequential sampling plan
together cause rigidity through an ``extreme'' overestimate of $\pi(d_2)$ based on
small sample size.
 The probability of this extreme overestimation is nonnegligible indeed:
if dose
level 2 is the true MTD with $\pi(d_2) = 0.20$, then the probability
that the trial is confined to the suboptimal dose 1 is at least
0.36 by a simple binomial calculation.
\citet{cheung02b} constructed a similar numerical example for the Bayesian
nonparametric curve-free method, and suggested that the rigidity
probability can be reduced by using an informative prior  to add
smoothness to the estimation.

Due to ethical constraints such as coherence and the discrete design space,
it may be challenging to achieve consistency without strong model assumptions.  For example,
the CRM has been shown to be consistent under certain model misspecifications, but is
not generally so \citep{shen96}.
  In this context, \citet{cheung02}  introduced the indifference interval
as a sensitivity measure of how close a design may approach $\nu$ on the probability scale:

\begin{Property}[(Indifference interval)]\label{pr3}  The indifference interval of a
 dose-finding design $\mathcal{D}$ exists and is equal to
$p \pm \delta$ if there exist $N > 0$ and $\delta \in (0,p)$ such that
\[
P_{\mathcal{D}} \{ x_n \in \mathcal{I}_{\pi}(p-\delta,p+\delta) \mbox{ for all $n \geq N$}
\}=1.
\]
\end{Property}

Apparently, the smaller the half-width $\delta$ of a design's indifference interval is, the
closer the design converges to the MTD; whereas a large $\delta$ indicates the design is
sensitive to the underlying $\pi$.
The sensitivity of the design $\mathcal{D}$ can thus be measured by $\delta$.
Specifically, a design with half-width $\delta$ (for some $\delta < p$) will be
called a $\delta$-sensitive design.

It is clear that if a design $\mathcal{D}$ is consistent for $\nu$, then it is
$\delta$-sensitive; that is, one may choose $\delta$ so that $\pi(\nu) \in p \pm \delta$.
Also, if $\mathcal{D}$ is  $\delta$-sensitive, then it is nonrigid.
Thus, while consistency appears to be too difficult and nonrigidity too
nondiscriminatory for a dose-finding design, $\delta$-sensitivity
seems to be a reasonable design property.
\citet{cheung02} prescribed a way to
calculate the indifference interval of the CRM, that is, the CRM is $\delta$-sensitive.
 Moreover,
\citet{lee09} showed that the CRM can be calibrated to achieve
any $\delta$ level of sensitivity.  However, it should be  noted
 that indifference interval is an asymptotic
criterion.  As such, a small $\delta$ does not necessarily yield good finite-sample properties.

\subsubsection{Unbiasedness}\label{sec:unbiased}
The performance of a reasonable dose-finding design is expected to improve
as the underlying dose-toxicity curve $\pi$ becomes steep.
This property, called unbiasedness by \citet{cheung07}, is formulated as follows:

\begin{Property}[(Unbiasedness)]\label{pr4}  Let $p_i = \pi(d_i)$ denote the true
toxicity probability at dose $d_i$.  A design $\mathcal{D}$ is said to be
unbiased if
\begin{longlist}[(a)]
\item[(a)] $P_{\mathcal{D}}( x_n = d_k)$ is nonincreasing in $p_{i'}$ for $i' \leq k$, and

\item[(b)] $P_{\mathcal{D}}(x_n = d_k)$ is nondecreasing in $p_i$ for $i > k$.
\end{longlist}
\end{Property}

For the special case with $d_k=\nu$ and $\pi(\nu) = p$, unbiasedness implies that the probability
of correctly selecting $\nu$ increases as the doses above the MTD become
more toxic (i.e., $p_i \gg p$), or the doses below  less toxic (i.e., $p_{i'} \ll p$).
In other words, the design will select the true MTD more often as it becomes
more separated from its neighboring doses in terms of toxicity probability.
A design that satisfies this special case is called {\em  weakly unbiased}.

\begin{figure}[t]

\includegraphics{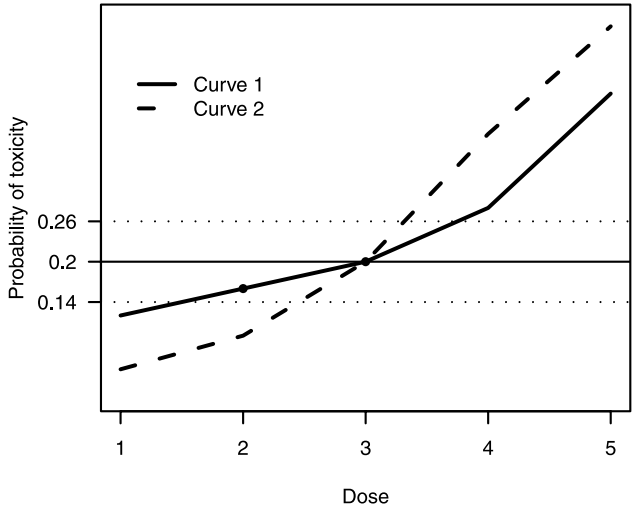}

\caption{Two dose-toxicity curves under which dose 3 is the MTD with $p=0.20$.
A $\delta$-sensitive design with $\delta =0.06$ will eventually select doses 2 or 3
under the shallow curve (curve 1), but will be consistent for dose 3 under the
steep curve (curve 2).  The horizontal dotted lines indicate the indifference interval.
} \label{fig:steep}
\end{figure}

One may argue that $\delta$-sensitive designs (e.g., the CRM)
are {\em asymptotically weakly unbiased}, in that they will be consistent
if the  underlying dose-toxicity curve $\pi$ becomes sufficiently steep around
the MTD; see Figure~\ref{fig:steep} for an illustration.
Unbiasedness has been established
 only for few designs in the dose-finding literature; an example
is the class of stepwise procedures \citep{cheung07}.  In practice,
extensive simulations are usually required, and are often adequate, to
confirm that a design is (weakly) unbiased.

\section{Stochastic Approximation}\label{sec:past}

\subsection{\texorpdfstring{The Robbins--Monro (\protect\citeyear{rm51})
Procedure}{The Robbins--Monro (1951) Procedure}}\label{sec3}

\citet{rm51} introduced the first formal stochastic approximation procedure
for the problem of finding
the root of a regression function.  Precisely,
let $M(x)$ be the mean of an outcome variable $Y=Y(x)$ at  level $x$, and suppose
$M(x) = \alpha$ has a unique root $\theta$ and
$\sup_x E\{ Y^2(x) \} < \infty$.
  Then the stochastic approximation recursion approaches
$\theta$ sequentially:
\begin{equation}\label{eq:sa}
x_{i+1} = x_i - \frac{1}{ib}(Y_i - \alpha)
\end{equation}
for some constant $b >0$.  It is well established that $x_n \rightarrow \theta$ with probability 1.
If in addition, the constant $b$ is chosen properly namely $b < 2 M'(\theta) \equiv 2 \beta$, then
$n^{1/2}(x_n -\theta)$ will converge in distribution to a normal variable with mean 0 and
variance $\sigma^2\{b(2\beta - b)\}^{-1}$ where
$\sigma^2 = \lim_{x \rightarrow \theta} \operatorname{var}\{ Y(x)\}$; see the works of \citet{sacks} and \citet{wasan}.

It is immediately clear that (\ref{eq:sa}) is  applicable to address objective
(\ref{eq:obj0}) in a clinical trial setting
with $M = \pi$ and
$\alpha = p$.  For one thing, the recursion output is  coherent (Property~\ref{pr1}) thus passing the first
ethical litmus test.
It is also easy to see that
a small-group sequential version of (\ref{eq:sa}), that is, replace $Y_i$ with $\bar Y_{i}$,
 is group coherent (Property~\ref{pr1'}).
There are, however, several practical considerations.

The choice of $b$ is crucial.  In view of efficiency, the asymptotic variance is minimized when we set
$b = \beta$, which is typically unknown in most applications.  This leads to the idea of
adaptive stochastic approximation where $b$ is replaced by a sequence $b_i$ that is strongly
consistent for $\beta$ \citep{lai79}.  However, when the sample size is small-to-moderate, the numerical
instability induced by the adaptive choice $b_i$ may offset its asymptotic advantage.
In this article, for a reason described in Section~\ref{sec:virtual}, we  assume that a good choice of $b$
is available.

The next  practical issue is that (\ref{eq:sa}) entails the availability of a continuum of doses.
This is seldom feasible in practice.  In drug trials, dose availability
is often limited by the
dosage of a tablet.  For treatments involving combination of drugs administered multiple times
over a fixed period, each subsequent dose may involve increasing doses and/or frequency of
different drugs.  For example, Table~\ref{ta:velcade}
describes the dose schedules of bortezomib used in a dose-finding trial in patients
with lymphoma \citep{leonard}.
The first three  levels prescribe bortezomib at a fixed dose 0.7 mg/m$^2$
with increasing frequency, whereas the next two increments apply the same frequency
with increasing bortezomib doses.  While we are certain that the risk for toxicity
increases over each level, there is no natural scale of dosage (e.g., mg/m$^2$).
Thus, assuming that the toxicity probability $\pi(x)$ is well defined on a continuous
range
of $x$ is artificial.

\begin{table}[t]
\caption{Dose schedules of bortezomib used in Leonard et al. (2005)}
\label{ta:velcade}
\begin{tabular*}{\tablewidth}{@{\extracolsep{4in minus 4in}}ll@{}}
\hline
\textbf{Level} & \multicolumn{1}{c@{}}{\textbf{Dose and schedule within cycle}} \\
\hline
1 & 0.7 mg/m$^2$ on day 1 of each cycle  \\
2 & 0.7 mg/m$^2$ on days 1 and 8 of each cycle  \\
3 & 0.7 mg/m$^2$ on days 1 and 4 of each cycle  \\
4 & 1.0 mg/m$^2$ on days 1 and 4 of each cycle  \\
5 & 1.3 mg/m$^2$ on days 1 and 4 of each cycle  \\
\hline
\end{tabular*}
\end{table}

To tailor the stochastic approximation for the discrete objective $\nu$,
an obvious approach is to round the output of (\ref{eq:sa})
to its closest dose at each iteration.  For example,
suppose that the dose labels are $\{ 1, \ldots, K\}$, that is,
$d_k = k$. Then a discretized stochastic approximation may be expressed as
\begin{equation}\label{eq:dsa}
x_{i+1} = C \biggl\{ x_i - \frac{1}{ib} (\bar Y_i - p) \biggr\},
\end{equation}
where $C(x)$ is the rounded value of $x$ if $0.5 \leq x < K + 0.5$, and is set equal to 1 and $K$
respectively if $x < 0.5$ or $> K+0.5$.  Unfortunately, the discretized stochastic approximation is rigid
(Property~\ref{pr2}).  To illustrate, consider
applying (\ref{eq:dsa}) with $b=0.2$ and a target $p =0.20$ in a trial with $x_1=1$ and $m=2$.
  Then no toxicity event in the first group, that is,
$\bar Y_1=0$, gives $x_2 = 2$.  Further suppose that the second group has a 50\% toxicity rate ($\bar Y_2=0.5$).
This will bring the trial back to $x_3=C( 2 - 0.75) = 1$; it is easy to see that the remaining subjects will
receive dose 1.  To see how rigidity occurs for a general variable type, we observe
that since $x_i$ is an integer, the update $x_{i+1}$ according to (\ref{eq:dsa}) will
stay the same as $x_i$ if $ | (ib)^{-1} (\bar Y_i - p) | < 0.5$, whose probability approaches 1
at a rate of $O(i^{-2})$ according to Chebyshev's inequality if $Y_i$ has a finite variance.
If $Y_i$ is bounded (e.g., binary), the term
$C \{ (ib)^{-1} (\bar Y_i - p) \}$ will always be zero as $i$ becomes sufficiently large,
and will not contribute to future updates.  This problem, called
{\em discrete barrier}, is thus built by rounding and the fact that the design points take on a discrete
set of levels.
In the context of the CRM, \citet{shen96} pointed out similar difficulties in establishing the theoretical
properties of dose-finding methods due to the discrete barrier.
This is where the modern dose-finding literature departs from the
elegant stochastic approximation approach.

\subsection{Stochastic Approximation and Model-Based Methods}

The Robbins--Monro stochastic approximation is a
nonparametric procedure in that the
convergence results depend only very weakly on the true underlying $M(x)$.
For the case of normal $Y$, interestingly, \citet{lai79} showed that the recursion output in (\ref{eq:sa})
is identical to the solution $\tilde x_{i+1}$ of
\begin{equation}\label{eq:ls}
\sum_{j=1}^i Y_j - \{ \alpha + b (\tilde x_j - \tilde x_{i+1}) \} = 0
\end{equation}
which amounts to maximum likelihood estimation of $\theta$ under a simple linear regression model.
This connection between the stochastic approximation and a model-based approach motivates
the study of the maximum likelihood recursion in the work of
\citet{wu85}, \citet{wu86}, and
\citet{ying97} for data arising from the exponential family.  In particular, for binary  $Y$,
\citet{wu85} proposed the logit-MLE that uses the logistic working model
\begin{eqnarray}\label{eq:logit}
F(x,\theta) = \frac{p \exp \{  \tilde b (x - \theta) \}}{1-p + p \exp \{ \tilde b (x - \theta)
\}}\nonumber
\\[-8pt]\\[-8pt]
  \eqntext{\mbox{for some $\tilde b >0$}}\vspace*{6pt}
\end{eqnarray}
and replaces the estimating equation (\ref{eq:ls}) with\break
$\sum_{j=1}^i \{ Y_j - F(\tilde x_j, \tilde x_{i+1}) \}=0$.
Here, we focus on the nonadaptive version, that is,
where $\tilde b$ is a fixed constant.
 A maximum likelihood version
 of the CRM (\ref{eq:crm})  would clearly yield the same design point as $\tilde x_{i+1}$
if the design space was continuous.  In this regard, the likelihood CRM
is an analogue of the logit-MLE for the discrete objective~$\nu$.

In order to establish the asymptotic distribution of the
logit-MLE (and the maximum likelihood recursion in general),
\citet{ying97} showed that the sequence $\tilde x_{i+1}$ is asymptotically equivalent to
an adaptive Robbins--Monro recursion; see the proof of Theorem 3 in the article by \citet{ying97}.
While the justification of the model-based logit-MLE relies on its asymptotic equivalence to
the nonparametric Robbins--Monro procedure, \citet{wu85} showed by simulation
 that
the former is superior to the latter in finite-sample settings with binary data.
Similarly, \citet{oq91} demonstrated that the CRM performs better than
 the discretized stochastic approximation (\ref{eq:dsa}) for the objective $\nu$.

These observations regarding the stochastic approximation, the logit-MLE, and the CRM bear
two practical suggestions.  First, in typical dose-finding trial settings
 with binary data and small sample sizes, a~model-based approach seems to retain some
information that is otherwise lost when using nonparametric procedures.
This speculation is made\break without assuming much confidence about the working model.
Second, one may study the theoretical (asymptotic) properties of the modern model-based method (e.g., CRM)
 by tapping the rich stochastic approximation literature, thus giving guidance on the choice
of design parameters such as $\tilde b$ in (\ref{eq:logit}).  This can be achieved, of course,
only
if we can resolve the discrete barrier---we will return to this in
 Section~\ref{sec:virtual}.

\subsection{Stochastic Approximation and Adaptive Control}

Maximum likelihood recursion attempts to optimize the prospect for the next subject by
setting the next design point at the current estimate of $\theta$, and is  myopic in that it does not consider the dose assignments of future subjects.
The Robbins--Monro procedure is therefore myopic by (asymptotic) equivalence.
\citet{lai79} studied the adaptive cost control aspect of the stochastic approximation
for normal $Y$ where
$C_n = \sum_{i=1}^n(x_i - \theta)^2$ is defined as
 the cost of a design sequence $\{x_i\}$ at stage $n$.
Specifically, they showed that the cost of (\ref{eq:sa}) is of the order $\sigma^2  \log n$ if
$b < 2 \beta$.  Under some simple linear regression models, \citet{han06} showed that the myopic
Bayesian rule is optimal when the slope parameter is known.
This suggests that the myopic Robbins--Monro method may also have good adaptive control properties.

The control aspect of the stochastic approximation is less clear for binary data.
\citet{bartroff} addressed the control problem by   using  techniques in
approximate dynamic programming  to minimize some well-defined global risk, such as
 the expectation of the design cost $C_n$.
 The authors demonstrated reduction of the global risk by
 nonmyopic approaches when compared to the myopic  ones including
the stochastic approximation and the logit-MLE.
The scope of the simulations, however, is confined
to situations where the logistic model correctly specifies $\pi$.
In addition, their approach is intended for the continuous objective $\theta$,
instead of $\nu$.

Further research on the use of nonmyopic\break approaches
in dose-finding is warranted, especially
 for practical situations with a discrete set of test doses.  The design cost at stage $n$ for
the discrete objective $\nu$ can be analogously defined as
$ C_n' =\break \sum_{i=1}^n (x_i - \nu)^2$.  Then a dose-finding design $\mathcal{D}$ is consistent if and only if
$\lim_{n \rightarrow \infty} C_n'$ is  finite almost everywhere.
As mentioned earlier, the myopic CRM is not necessarily consistent (as it tries to treat each subject
at the current ``best'' dose).  By contrast, designs that spread out the design points
(e.g., the biased coin design) allow consistent estimation of $\nu$ at the expense of the enrolled
subjects.  Neither guarantees a finite $\lim_n C_n'$.
An optimal $\mathcal{D}$ for the infinite-horizon control of $C_n'$ thus seems to  resolve
 the inherent tension between the welfare of enrolled subjects (i.e., the cost is kept low)
and the estimation of $\nu$ (i.e., $x_n$ is consistent).

\section{Ongoing Relevance}\label{sec:practice}

\subsection{Binary Versus Dichotomized Data}

As mentioned earlier, with a binary outcome and small samples,
the Robbins--Monro procedure is generally
less efficient than model-based methods,
and hence may not be suitable for clinical dose-finding where the study
endpoint is classified as toxic and nontoxic.  In many situations, however, the binary
toxic outcome $T$ is defined by dichotomizing an observable biomarker expression $Y$, namely,
$T = \mathbf{1}( Y > t_0)$ for some fixed safety threshold $t_0$, where $\mathbf{1}(E)$
denotes the indicator of the event $E$.  The biomarker $Y$ apparently contains more information
than the dichotomized $T$, and may be used to achieve the dose-finding objective~(\ref{eq:obj0})
with greater efficiency.

To illustrate,
consider the regression model
\begin{equation}\label{eq:loc-scale}
Y = M(x) + \sigma(x) \epsilon,
\end{equation}
where $\epsilon$ has a known distribution $G$ with mean 0 and variance 1.  Under
(\ref{eq:loc-scale}), the toxicity probability can be expressed as
$\pi (x) = 1 - G [ \{ t_0 - M(x)\}/ \sigma(x) ]$
and the continuous dose-finding
objective (\ref{eq:obj0}) can be shown to be equivalent to the solution
to
\begin{equation}\label{eq:obj-c}
f(x) \equiv M(x) + z_p \sigma(x) = t_0,
\end{equation}
where $z_p$ is the upper $p$th percentile of $G$.
To focus on the comparison between the use of $Y$ and $T$, suppose for the moment
that  a continuum of  dose $x$ is available.
Further suppose that  a trial enrolls patients in small groups of size $m$.  Let $x_i$ denote the
dose given to the $i$th group, and $Y_{ij}$ the biomarker expression of the $j$th subject in the group.
With this experimental setup, we note that
\begin{equation}\label{eq:unbiasO}
O_i = \bar{Y}_i + [ E \{ S_i / \sigma(x_i) \} ]^{-1} z_p S_i
\end{equation}
is an unbiased realization of $f(x_i)$, where $S_i$ is the sample standard deviation of the observations
in\break group~$i$.  The expectation in (\ref{eq:unbiasO}) can be computed for any given $G$, because
$S_i/\sigma(x_i)$ depends on the error variable $\epsilon$
 but not $M$ and $\sigma$ under model (\ref{eq:loc-scale}).
In other words, $O_i$ is observable and is a continuous variable that can be used to generate a stochastic
approximation recursion
\begin{equation}\label{eq:osa}
x_{i+1} = x_i - (ib)^{-1} (O_i - t_0).
\end{equation}
The design $\{ x_n \}$ generated by (\ref{eq:osa}) is consistent for $\theta$ under the condition that
$\theta$ is the unique solution to (\ref{eq:obj-c}).  This condition holds, for example, when $M$ is
strictly increasing
and $\sigma$ is nondecreasing in $x$.  This
is a reasonable assumption for many biological measurements,
for which the variability typically increases with the mean.
Furthermore, if $b < 2 \beta$, where $\beta = f'(\theta)$ here, then  the asymptotic variance of $x_n$ is
$v_O = \lim_{x \rightarrow \theta} \operatorname{var}(O_i)  \{ b (2\beta - b) \}^{-1}$.
In particular, when $\epsilon$ is standard normal,
\[
v_O = \frac{ \sigma^2(\theta) \{ 1+ m z_p^2 (\lambda_m -1) \}}{mb(2\beta -b)},
\]
where
\[
\lambda_m = \frac{(m-1) \Gamma^2\{ (m-1)/2 \}}{2 \Gamma^2(m/2)}.
\]

Now, instead of using the recursion (\ref{eq:osa}), suppose that we apply the logit-MLE based on the\break
dichotomized outcomes by solving
$\sum_{j=1}^i \{ \bar{T}_j -\break F(\tilde x_j, \tilde x_{i+1} )\} = 0$
where $F$ is defined in (\ref{eq:logit}).  Then using the results in the article of \citet{ying97}, we can show that
$\sqrt{n} (\tilde{x}_n - \theta)$ converges in distribution to a mean zero normal with
variance $v_T = p(1-p) \{ m \tilde{b} (2 \tilde \beta - \tilde b) \}^{-1}$ where $\tilde \beta \equiv \pi'(\theta) =
\beta G'(z_p) / \sigma(\theta)$.

\begin{figure}[t]

\includegraphics{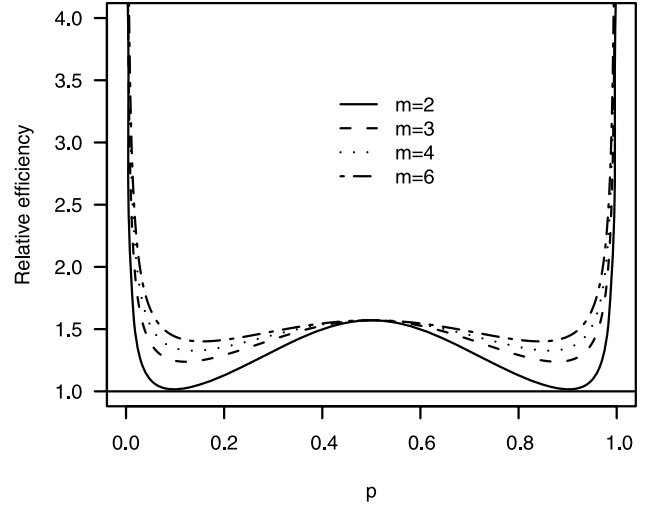}

\caption{Asymptotic efficiency of $x_n$ based on
recursion (\protect\ref{eq:osa}) relative to the logit-MLE $\tilde x_n$.}
\label{fig:eff}
\end{figure}

The asymptotic variances of $v_O$ and $v_T$ are minimized when
$b = \beta$ and $\tilde b = \tilde \beta$, respectively.  Thus, the optimal choice depends on
unknown parameters.  For the purpose of comparing efficiencies,
suppose we could set $b$ and $\tilde{b}$ to their respective
optimal values.  Then the variance ratio is equal to
\begin{equation}
\frac{v_T}{v_O} = \frac{p(1-p)}{ \{ G'(z_p) \}^2 \{ 1 + m z_p^2 (\lambda_m - 1) \}}
\label{eq:eff}
\end{equation}
for normal noise, and  also
represents the asymptotic efficiency of $x_n$ relative to $\tilde x_n$.
For $m=3$, the ratio (\ref{eq:eff})
 attains a minimum of 1.238 when $p=0.12$ or 0.88.
 As shown in Figure~\ref{fig:eff}, the efficiency gain can be substantial for
 any group sizes larger than 2, especially when the target $p$ is extreme.

\subsection{Virtual Observations}\label{sec:virtual}

A particular
obstacle to the use of stochastic approximation is the discrete
 design space used in clinical studies, which creates the discrete barrier (Section~\ref{sec3}).
To overcome the discrete barrier,  \citet{cheung10} introduced the notion of virtual observations.
Precisely, the virtual observation of the $i$th group of subjects
is defined as
\begin{equation}\label{eq:vo}
V_i = O_i + b (x_i^{*} - x_i),
\end{equation}
where $x_i^{*}$ denotes the {\em assigned} dose of the group which can take values on a
 continuous conceptual
scale that represents an ordering of doses.
In the situations where the {\em actual given} dose $x_i$ can take on any real value, we
have $x_i^{*} \equiv x_i$
and $V_i \equiv O_i$, and thus, the recursion (\ref{eq:osa}) may be used to approach the target dose $\theta$.
When $x_i$ is confined to $\{1, \ldots, K\}$, \citet{cheung10} proposed generating a stochastic approximation
recursion based on the virtual observations:
\begin{equation}\label{eq:savo}
x_{i+1}^{*} = x_i^{*} - \frac{1}{ib} (V_i - t_0),
\end{equation}
and treating the next group of subjects at
$x_{i+1} = C(x_{i+1}^{*})$.  To initiate the virtual observation recursion,
one may set $x_1^{*} = x_1 \in \{1,\ldots, K\}$.

\citet{cheung10} proved, under mild conditions, that $x_{i+1}^{*}$ generated by (\ref{eq:savo}) is consistent
(hence  nonrigid) for $\theta_b$ for some $\theta_b = \nu \pm 0.5$, and hence $x_{i+1}$ for $\nu$.
 Briefly, for any given $b$,
consistency will occur if the neighboring doses of the MTD are sufficiently apart from the MTD in terms of toxicity probability.
This is in essence asymptotically weakly unbiased as defined in Section~\ref{sec:unbiased}, and can be easily
derived from Propositions 2 and 3 of \citet{cheung10}.

With the use of continuous $V$'s,
the notion of coherence needs to be re-examined.  In particular, the
virtual observation recursion (\ref{eq:savo}) will de-escalate if the biomarker expression of the current subjects
has a high average ($\bar Y_i$) or a large variability ($S_i$).  This is a sensible dose-escalation principle for situations where
the variability increases proportionally to the mean.

The idea of virtual observation is to create an objective function
\[
h(x) = E(V_i | x_i^{*} =x) = f\{ C(x) \} + b \{ x - C(x) \}
\]
that is defined on the real line, and has a local slope at $\{1, \ldots, K\}$, such that the solution $\theta$ of
(\ref{eq:obj-c}) can be approximated by the solution $\theta_b$ of $h(x) = t_0$.  Quite
importantly, since now the objective function $h$
has a known slope $b$ around $\theta_b$ (under some  Lipschitz-type regularity conditions), we can use
the same  $b$ in the recursion (\ref{eq:savo}) as in the definition of virtual observations (\ref{eq:vo}).
This design feature enables us to achieve optimal asymptotic variance without resorting to
adaptive estimation of the slope of the objective function.  It is particularly relevant to early phase dose-finding
studies where adaptive stochastic approximation can be unstable due to small sample sizes.

\section{Looking to the Future}\label{sec:future}

Statistical methodology for dose-finding trials is by its
nature an application-oriented discipline.
Consequently, much of the emphasis in the dose-finding
literature has been on empirical properties via simulation.
 However,
as the (model-based) methods become increasingly complicated, it is imperative to check their properties
against some theoretical criteria so as to avoid pathological behaviors
that may not be detected in aggregate via simulations; rather, pathologies such as incoherence and rigidity
are {\em pointwise} properties that can be found by careful analytical study.
As a case in point, the virtual observation recursion
 (\ref{eq:savo}) is presented in light of the properties described in
Section~\ref{sec:theory}.
Granted, as the data content becomes richer, these theoretical criteria have to be re-examined.
\citet{cheung-crc}, in another instance,
 extended  the notion of coherence for bivariate dose-finding in the context of  phase I/II trials---see the article by
\citet{thall-sts} for a review of the bivariate dose-finding objective---and showed how coherence can
 be used to simplify dose decisions in the complex ``black-box'' approach of the bivariate model-based
methods, and to provide clinically sensible rules.

The idea of virtual observation  bridges
the stochastic approximation
and the modern (model-based) dose-finding literatures.
As the Robbins--Monro\break method has motivated a large number of extensions and refinements for a wide
variety of  root-finding objectives, there exists a
reservoir of ideas from which we can borrow and apply to
dose-finding methods for specialized clinical situations.
To name a few,  consult the works of
\citet{kw52} for finding the maximum of a regression function, and \citet{blum54} for multivariate contour-finding.
While studying the analytical properties of model-based designs in these specialized situations can be difficult,
connection to the theory-rich stochastic approximation procedures  allows us to do so with relative ease
and elegance, as is the case for the virtual observation
recursion (\ref{eq:savo}).  In this light,  extending the idea
of virtual observations for  data types other than continuous and multivariate data appears to be a promising
``crosswalk'' that warrants further research.

\section*{Acknowledgment}
This work was supported by NIH Grant R01 NS0558 09.

\end{document}